\documentclass[sigconf]{acmart}

\usepackage{pifont}

\definecolor{keyword}{HTML}{2F7AFF}
\definecolor{identifier}{HTML}{000000}
\definecolor{commentstyle}{HTML}{5E5E5E}
\definecolor{cell1}{HTML}{FF5B00}
\definecolor{cell2}{HTML}{ffa200}
\definecolor{cell3}{HTML}{f9da24}
\definecolor{cell4}{HTML}{f9da24}
\definecolor{cell5}{HTML}{cff09e}
\usepackage{colortbl}

\usepackage{enumitem}
\usepackage{subfig}
\usepackage{multirow}
\usepackage{makecell}

\usepackage{listings}
\lstdefinelanguage{mylanguage}{
  keywords={True, False, return, switch, if, in, while, do, else, case, break, def, for},
  keywordstyle=\color{keyword}\bfseries,
  ndkeywords={class, export, boolean, throw, implements, import, this},
  ndkeywordstyle=\color{darkgray}\bfseries,
  identifierstyle=\color{identifier},
  sensitive=false,
  comment=[l]{\#},
  morecomment=[s]{/*}{*/},
  commentstyle=\color{commentstyle}\ttfamily,
  stringstyle=\color{red}\ttfamily,
  morestring=[b]',
  morestring=[b]"
}
\lstdefinestyle{mystyle}{
  language=mylanguage,
  backgroundcolor=\color{white},
  numberstyle=\tiny\color{black},
  basicstyle=\ttfamily\footnotesize,
  breakatwhitespace=false,
  breaklines=true,
  captionpos=b,
  keepspaces=true,
  numbers=left,
  numbersep=3pt,
  stepnumber=1,
  showspaces=false,
  showstringspaces=false,
  showtabs=false,
  tabsize=2,
  morekeywords={entry},
  frame=tb,
}
\usepackage[ruled,vlined,linesnumbered]{algorithm2e}

\usepackage{amsfonts}
\DeclareMathOperator*{\argmax}{arg\,max}

\usepackage{xspace}
\usepackage{soul}
\usepackage{lscape}

\makeatletter
\renewcommand{\@specialsection}[1]{%
  \let\@vspace\@vspace@orig
  \let\@vspacer\@vspacer@orig
  \par\medskip\small\noindent\textbf{\textit{#1:}}\enspace}
\makeatother

\newcommand{\compactparagraph}[1]{\noindent{\textbf{\textit{#1}}}}
\newif\ifcrmarked
\crmarkedfalse
\newcommand{\crcolor}{}
\newcommand{\cradd}[1]{{#1}}
\newenvironment{crrevision}{}{}
\newcommand{\crplatformrate}{5}
\newcommand{\revisioncr}[1]{#1}

\newcommand{\projectname}{{\tt EAServe}\xspace}
\newcommand{\projectnamenott}{EAServe\xspace}

\copyrightyear{2026}
\acmYear{2026}
\setcopyright{cc}
\setcctype{by}
\acmConference[PACT '26]{International Conference on Parallel Architectures and Compilation Techniques}{October 19--22, 2026}{Chicago, IL, USA}
\acmBooktitle{International Conference on Parallel Architectures and Compilation Techniques (PACT '26), October 19--22, 2026, Chicago, IL, USA}
\acmDOI{10.1145/3838684.3846886}
\acmISBN{979-8-4007-2915-7/2026/10}
\hypersetup{keeppdfinfo}
\begin{document}

\title{\projectnamenott: Encode-Aware Disaggregated Serving for Multimodal Large Language Models}
\renewcommand{\shorttitle}{\projectnamenott}

\author{Kunxiong Zhu}
\email{Kunxiong.Zhu@uga.edu}
\affiliation{%
  \institution{University of Georgia}
  \city{Athens}
  \state{GA}
  \country{USA}
}
\author{Zhihao Shu}
\email{Zhihao.Shu@uga.edu}
\affiliation{%
  \institution{University of Georgia}
  \city{Athens}
  \state{GA}
  \country{USA}
}
\author{Hangyu Zheng}
\email{hyzheng@uga.edu}
\affiliation{%
  \institution{University of Georgia}
  \city{Athens}
  \state{GA}
  \country{USA}
}
\author{Minghai Qin}
\email{qinminghai@gmail.com}
\affiliation{%
  \institution{Western Digital}
  \city{San Jose}
  \state{CA}
  \country{USA}
}
\author{Miao Yin}
\email{miao.yin@uta.edu}
\affiliation{%
  \institution{University of Texas at Arlington}
  \city{Arlington}
  \state{TX}
  \country{USA}
}

\author{Gagan Agrawal}
\email{gagrawal@uga.edu}
\affiliation{%
  \institution{University of Georgia}
  \city{Athens}
  \state{GA}
  \country{USA}
}

\author{Wei Niu}
\email{wniu@uga.edu}
\affiliation{%
  \institution{University of Georgia}
  \city{Athens}
  \state{GA}
  \country{USA}
}

\begin{abstract}

With growing emphasis on (text-only) LLM serving, 
disaggregating  the two stages: \textsc{Prefill} and \textsc{Decode} onto separate GPU pools
is now a  standard optimization. 
However, multimodal LLMs (MLLMs),  which  add a third phase, \textsc{Encode}, pose new 
challenges for optimizing resource allocation. 
\textsc{Encode} turns images, video, or audio into embeddings  that the language model can consume,
yielding a three-stage Encode--Prefill--Decode (EPD) pipeline, which requires new research for 
optimizing resource allocation and utilization. Existing frameworks offer only partial answers:
text-only PD systems lack \textsc{Encode}, while EPD frameworks expose it as a separate service
without regulating downstream request flow.
New challenges arise because the three stage pipeline carries a structural resource imbalance:
every request enters through \textsc{Encode} before downstream work can begin,
yet per-request execution leaves the encode GPU severely underutilized even at high loads,
and this idle capacity starves the downstream \textsc{Prefill} and \textsc{Decode} workers.

Addressing this challenge, 
we  reposition \textsc{Encode} as the control point of the EPD pipeline,
exposing three tightly coupled dimensions for optimization: 
when work enters downstream, where prefill executes, and how the GPU is shared.
We instantiate this in EAServe across two co-designed layers.
Its runtime manages load-adaptive micro-batching, rate-controlled partial offload to a co-resident prefill worker,
and dynamic SM partitioning for predictable co-location.
The configuration layer, Hybrid Auto Selection (HAS), then navigates the joint space of GPU allocation, 
encode batch size, and offload ratio by pruning unbalanced allocations 
with per-stage capacity profiling and refining the remainder through TPE-based Bayesian optimization. 
In our extensive evaluation on three MLLM architectures spanning image, video, and audio, EAServe delivers up to 4.3$\times$ and 1.7$\times$ higher goodput than NVIDIA Dynamo and vLLM, respectively, under identical SLO constraints,
sustains substantially more balanced and higher GPU utilization across the EPD pipeline,
while reaching near-optimal disaggregated configurations faster than baseline search methods.
\end{abstract}



\maketitle

\section{Introduction}
\label{sec:introduction}

Multimodal Large Language Models (MLLMs) are rapidly moving
from research prototypes to deployed services, 
as  they now power interactive assistants, visual question answering,
document understanding, video understanding, and speech-based applications~\cite{man2025adacm,liu2023llava,bai2023qwenvl,li2024llavaonevision,fixie2024ultravox}.
As deployment grows, the systems challenge is shifting from supporting MLLMs at all
to serving them efficiently at scale under latency objectives and fluctuating demand~\cite{hydrainfer2025}.

Serving an MLLM efficiently is more challenging than serving a text-only LLM
because the pipeline is more heterogeneous.
Text-only inference splits into a compute-bound \textsc{Prefill} 
and a memory-bandwidth-bound \textsc{Decode},  or PD, 
and this clean split is what makes {\em disaggregating}  the two onto independently scaled 
GPU pools ({\em PD separation}) effective in practice~\cite{zhong2024distserve,patel2024splitwise}.
MLLMs add a third stage in front of this pipeline. Specifically,
a  \emph{modality encoder}, typically a  transformer, turns images, video,
or audio into embeddings that the language model can then consume.  
The result is  the
encode-prefill-decode (EPD) pipeline, which is widely  used in modern MLLM
deployments~\cite{nvidia2025dynamo}.
The new \textsc{Encode} stage's 
compute and bandwidth demands vary with modality, input size, and arrival pattern.
This extra axis of heterogeneity makes EPD harder to schedule than PD.

Despite being  on the critical path, \textsc{Encode} has received less attention
from recent work than the two stages  following  it.
Text-only PD frameworks~\cite{zhong2024distserve,patel2024splitwise}
expose only a token-level interface and coordinate \textsc{Prefill} with \textsc{Decode}.
EPD frameworks such as NVIDIA Dynamo~\cite{nvidia2025dynamo}
factor \textsc{Encode} out as a separate service but operate it as a pass-through:
one request at a time, with operator-set GPU allocation and no mechanism for regulating
how work enters the rest of the pipeline.
The cost of this passivity is that
a single encoder forward pass is too small to saturate a modern GPU,
yet batching is constrained by the latency cost of waiting for a larger batch to form.
The stage that gates pipeline admission is therefore also the one that leaves the most capacity unused:
under sustained load, the encode GPU stays largely idle 
while downstream workers wait on a growing encode queue and tail latency inflates~\cite{nvidia2025dynamo}.

We argue that \textsc{Encode} should instead be repositioned as the pipeline's \emph{control point}.
As the pipeline's entry, it naturally controls three coupled dimensions:
\emph{when} work enters downstream, \emph{where} prefill executes,
and \emph{how} the GPU is shared under co-location.
Co-optimization is needed as tuning any of them in isolation breaks the others, 
and yet challenging, as  joint space is too large to search by hand. 
With this motivation, this paper presents \projectname, a layered serving architecture built around this control point.
At the top, Hybrid Auto Selection (HAS) performs offline configuration search over the
joint space of GPU allocation, encode batch bound, and prefill offload ratio.
At the bottom, three encode-aware runtime mechanisms expose the parameters that HAS
searches and enforce the latency guarantees that make the search results reliable.
Different mechanisms dominate on different modalities, 
motivating both the runtime co-design and the joint search.

In all, 
this paper makes the following contributions.
\begin{itemize}[leftmargin=*,noitemsep,nolistsep]
    \item
    We identify \textsc{Encode} as the natural control point
    of disaggregated MLLM serving: both the entry point of the pipeline
    and a major source of unused GPU capacity.

    \item
    We design \emph{Hybrid Auto Selection} (HAS), an offline framework that
    automatically picks the deployment configuration of the EPD pipeline for
    a given serving workload.

    \item
    We design \emph{three encode-aware runtime mechanisms} that turn HAS's chosen configuration
    into a working serving stack and adapt to load online with no per-deployment tuning:
    load-adaptive micro-batching with a Poisson-gap dispatch rule,
    rate-controlled partial offload to a local prefill worker,
    and dynamic per-micro-batch Streaming Multiprocessor (SM) partitioning.

    \item
    We implement \projectname and show significant speedups
     over the state-of-the-art production-level baselines.
\end{itemize}

We evaluate \projectname on three representative MLLMs covering the main input
modalities: \textsc{LLaVA-v1.6-34B}~\cite{liu2023llava} for image, \textsc{Qwen2.5-VL-32B}~\cite{bai2025qwen25vl} for video,
and \textsc{Ultravox-v0.6-27B}~\cite{fixie2024ultravox} for audio.
Experiments run on an 8-GPU server under a range of TTFT/TPOT latency targets,
and we compare against the state-of-the-art baselines:
NVIDIA Dynamo~\cite{nvidia2025dynamo} and vLLM~\cite{kwon2023vllm}.
Across these workloads, \projectname improves goodput (SLO-compliant request rate)
by up to $4.3\times$ over Dynamo and $1.7\times$ over vLLM,
and yields more balanced GPU utilization across the EPD pipeline.

\section{Background and Motivation}
\label{sec:background}

\subsection{EPD Pipeline And Metrics}
\label{sec:bg-pipeline}

\begin{figure}[t]
  \centering
  \includegraphics[width=1\linewidth]{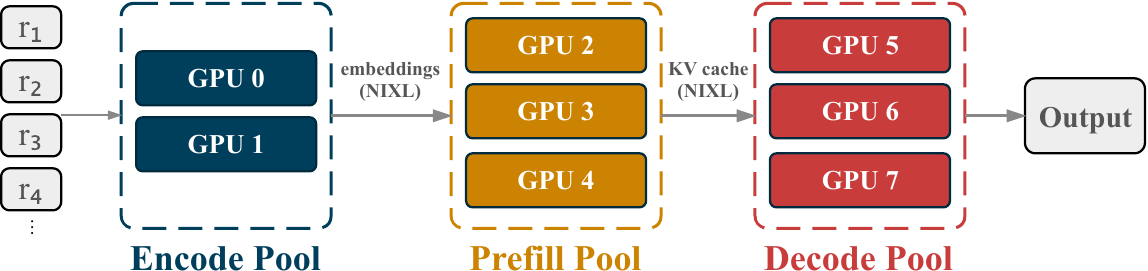}
  \vspace{-0.5em}
  \caption{The disaggregated pipeline of EPD.} %
  \Description{Block diagram of the three-stage EPD serving pipeline. Incoming requests enter an Encode Pool holding GPU 0 and GPU 1. Encoded embeddings are transferred over NIXL to a Prefill Pool holding GPU 2, GPU 3 and GPU 4. The resulting KV cache is transferred over NIXL to a Decode Pool holding GPU 5, GPU 6 and GPU 7, which produces the output. Each pool is a separately provisioned group of GPUs.}
  \label{fig:epd-pipeline}
\end{figure}

An MLLM extends a text-only LLM with non-text inputs such as images, video, or audio. 
Broadly, MLLMs include
a \emph{modality encoder} (e.g., CLIP ViT~\cite{radford2021clip} for images, Whisper~\cite{radford2023whisper} for audio)
that maps the raw input into embedding vectors, 
a lightweight \emph{connector} that  projects them into the LLM's input space, 
and the \emph{LLM backbone} that consumes them alongside text tokens to generate text autoregressively.
Disaggregated MLLM serving thus has  three stages 
that scale on independent GPU pools (Figure~\ref{fig:epd-pipeline}):
\textsc{Encode} runs the encoder and connector to produce \emph{modality embeddings};
\textsc{Prefill} executes the backbone on the concatenation of modality embeddings and text prompt, 
populates the key--value (KV) cache, and emits the first output token;
\textsc{Decode} generates the remaining tokens autoregressively from the context.

This decomposition  results in stages with  heterogeneous resource profiles: 
\textsc{Encode} performs short forward passes; 
\textsc{Prefill} is a single attention-heavy pass whose cost grows with context length; 
and \textsc{Decode} runs a long, memory-bandwidth-bound autoregressive loop.
Independent scaling lets each stage be provisioned independently~\cite{nvidia2025dynamo}, 
at the cost of carrying intermediate state (modality embeddings and KV cache) across workers.
Within a stage, when the model exceeds one GPU, 
weights are usually sharded with tensor parallelism (TP), the standard intra-replica strategy in latency-sensitive serving; we denote  $\mathrm{TP}{=}k$ for $k$-way sharding.
In all, we denote the resulting GPU allocation across the three stages by $(E, P, D)$, 
which, together with the choice to enable co-located workers,
specifies the deployment configuration for a given (model, workload, cluster) combination.

A configuration's quality cannot be judged by throughput alone, 
since the EPD pipeline is expected to meet user-facing latency targets at every stage.
Following prior LLM-serving work~\cite{zhong2024distserve,patel2024splitwise,kwon2023vllm}, 
we therefore use latency  and SLO-aware metrics.
\emph{Time-to-first-token (TTFT)} is the elapsed time from request arrival to the first output token; 
in an EPD pipeline it covers encode, embedding transfer, prefill, and any queueing along the path.
\emph{Time-per-output-token (TPOT)} is the mean interval between consecutive output tokens during \textsc{Decode}.
We report both at the mean and the 99th-percentile tail (P99).
Following DistServe~\cite{zhong2024distserve}, 
we summarize end-to-end performance with \emph{goodput}, the rate of requests that satisfy the SLO:
\begin{equation}
\mathrm{Goodput} = N_{\text{SLO-met}} / T,
\label{eq:goodput}
\end{equation}
where $N_{\text{SLO-met}}$ counts requests with $\mathrm{TTFT}\le T_f$ and $\mathrm{TPOT}\le T_p$
over a run of duration $T$.
We adopt goodput as the primary metric in our evaluation (\S\ref{sec:evaluation}),
since it captures both throughput and SLO attainment in a single number.

\begin{figure}[t!]
  \centering
  \includegraphics[width=0.92\linewidth]{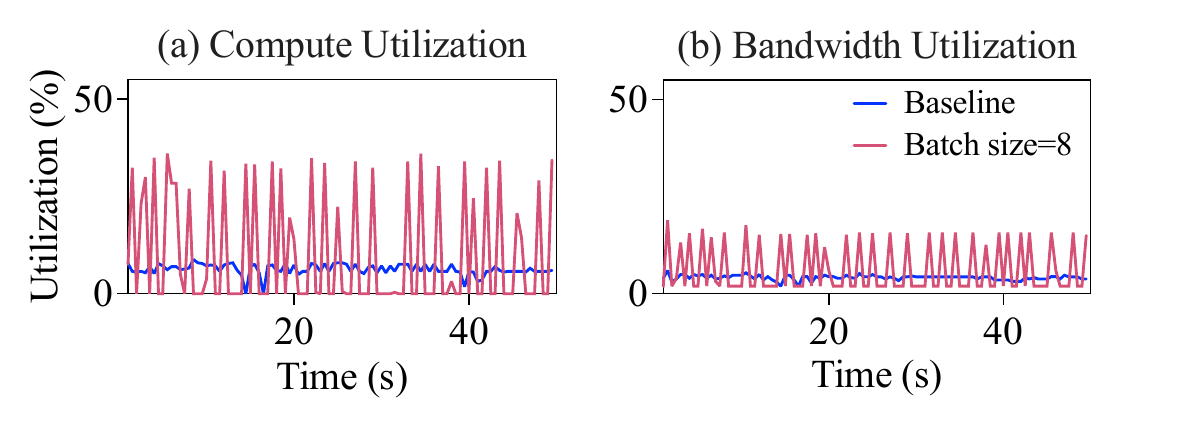}
  \vspace{-0.5em}
  \caption{GPU computation and HBM-bandwidth utilization on \textsc{LLaVA-v1.6-34B}~\cite{liu2023llava} under NVIDIA Dynamo~\cite{nvidia2025dynamo}.}%
  \Description{Two time-series plots covering a 50-second window on the encode GPU for LLaVA-v1.6-34B. Panel (a) plots compute utilization and panel (b) plots HBM bandwidth utilization, both on a vertical axis running from 0 to 50 percent. The baseline curve is nearly flat at roughly 5 percent in both panels. The batch-size-8 curve is spiky, peaking near 35 percent for compute and near 20 percent for bandwidth but falling back between spikes. Neither configuration comes close to saturating the encode GPU.}
  \label{fig:encode-utilization}
  \vspace{-0.5em}
\end{figure}

\subsection{Why Encode Becomes a  System Bottleneck}
\label{sec:mot}

As the entry point of the pipeline, \textsc{Encode} gates the  downstream stages; 
yet a single encode pass rarely places enough work on the GPU to keep it busy.
We illustrate the resulting inefficiency with two measurements taken on a standard EPD deployment
(Figure~\ref{fig:encode-utilization}). Full hardware, model, and workload details appear in \S\ref{sec:eval-setup}.

\compactparagraph{Observation 1. Encode gates the pipeline yet leaves substantial GPU capacity unused.}
Encode batching is constrained by request arrivals and the latency cost of waiting for a larger batch to form, 
so the system cannot grow the batch size at will.
Individual encoder passes are consequently too small to saturate the GPU: both compute and high-bandwidth memory (HBM) bandwidth utilization remain low (as shown in Figure~\ref{fig:encode-utilization}, Baseline).
At sustained arrival rates the per-request encoder throughput falls below the offered load,
so downstream \textsc{Prefill} and \textsc{Decode} workers spend most of their time waiting on a growing encode queue
rather than processing embeddings.

\compactparagraph{Observation 2. Micro-batching helps but does not close the gap.}
Adding a simple micro-batcher ($B{=}8$) raises throughput and lowers TTFT (Table~\ref{tab:motivation-e2e}), 
but mean TPOT rises sharply because the accelerated encode floods the downstream \textsc{Prefill} 
and \textsc{Decode} workers.
The encode GPU itself remains far from saturated, 
alternating between brief bursts and long idle gaps (Figure~\ref{fig:encode-utilization}, $B{=}8$).
Micro-batching captures part of the available gain but does not resolve the underlying imbalance.

\begin{table}[t]
  \centering
  \caption{A partial gain from adding micro-batching to Dynamo on \textsc{LLaVA-v1.6-34B}.}
  \setlength{\tabcolsep}{8pt}
  \label{tab:motivation-e2e}
  \small
  \begin{tabular}{l r r r}
    \toprule
    Configuration & Throughput & Mean TTFT & Mean TPOT \\
                  & (req/s)    & (s)       & (ms)      \\
    \midrule
    Dynamo default            & 0.75 & 367.4 & 179 \\
    \quad + micro-batch & 1.09 & 112.1 & 411 \\
    \bottomrule
  \end{tabular}
\end{table}

These two observations suggest three design choices for an encode-aware system.
First, a \emph{dispatch policy} must be decided  that adapts the encode batch bound to the
current arrival rate, so the dispatcher tracks load rather than waits on a static timeout.
Second, an \emph{offload mechanism} is needed that channels the encode-GPU's idle capacity
into a co-located prefill worker without slowing encode itself.
Third, a \emph{resource-sharing primitive} must keep the two co-located workers
isolated as the batch size changes.
The three need to be co-optimized: the batch bound shapes the SM split that
protects latency, the SM split shifts how much offload the local worker can
accept, and any reallocation across $(E, P, D)$ rebalances all three.
The resulting joint configuration is too large to tune by hand for each new
(model, workload, cluster), which is the challenge our work  addresses.

\begin{figure}[t!]
  \centering
  \includegraphics[width=0.8\linewidth]{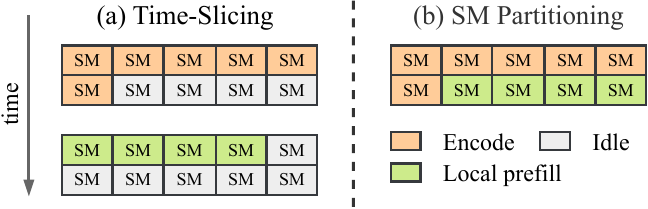}
  \vspace{-0.5em}
  \caption{SM partitioning via \texttt{libsmctrl}. (a) Default time-slicing: encode (orange) and local prefill (green) alternate, leaving idle SMs (gray). (b) TPC-mask partitioning runs them concurrently on disjoint SM subsets.}
  \Description{Schematic comparing two ways of sharing streaming multiprocessors between encode and local prefill, with time running downward. In panel (a), time-slicing, an all-orange encode row alternates with a row that mixes green local-prefill blocks and gray idle blocks, so only one kind of work runs at any instant. In panel (b), SM partitioning, orange encode blocks and green local-prefill blocks occupy disjoint SMs within the same row and run concurrently, leaving no idle blocks.}
  \label{fig:sm-partition}
\end{figure}

\begin{table}[t!]
  \centering
  \small
  \caption{A summary of key symbols.}
  \vspace{-0.5em}
  \label{tab:design-notation}
  \begin{tabular}{p{0.14\linewidth}p{0.78\linewidth}}
    \toprule
    Symbol & Meaning \\
    \midrule
    $PC = (\mathrm{alloc},\,B,\,s)$ & Deployment configuration searched by HAS, consisting of the stage allocation, maximum encode batch size, and offload ratio. \\
    $\mathrm{alloc}$ & GPU allocation across the \textsc{Encode}, \textsc{Prefill}, and \textsc{Decode} stages. \\
    $B$ & Maximum encode batch size. \\
    $s$ & Target fraction of requests routed to remote prefill. \\
    $\Lambda$ & Target system-wide arrival rate. \\
    $T_f,\; T_p$ & SLO thresholds on P99 TTFT and P99 TPOT, respectively. \\
    \bottomrule
  \end{tabular}
\end{table}

\subsection{SM Partitioning}
\label{sec:bg-mps}

A modern GPU offers hundreds of TFLOPS of compute and terabytes per second of HBM bandwidth, 
and this has motivated co-locating two workloads on one GPU for full utilization. 
Existing primitives form a hierarchy of increasing isolation strength, with implications
for how workloads can interfere with one another. 
The default CUDA scheduler simply time-slices the full SM array, 
so only one process runs at a time and concurrency is lost.
CUDA Multi-Process Service (MPS)~\cite{nvidia2020mps} merges kernels from multiple processes into a single CUDA context
to enable concurrent execution, optionally with per-client compute-percentage hints.  
However, 
the hints are advisory,  and a long-running kernel from one client 
can still delay latency-sensitive work in another, 
so this solution  raises throughput without protecting tail latency.
The strongest option is \emph{spatial partitioning}, 
which assigns disjoint physical resources to each process.
NVIDIA Multi-Instance GPU (MIG) does this in hardware on A100/H100-class GPUs 
and partitions both SMs and HBM bandwidth, 
but its reconfiguration is heavyweight and it is unavailable on the consumer 
and workstation GPUs widely used for MLLM serving.
Software-level \emph{spatial SM partitioning} via \texttt{libsmctrl}~\cite{libsmctrl} fills this gap
by writing per-process Texture Processing Cluster (TPC) masks (Figure~\ref{fig:sm-partition}): 
a process executes only on its allotted SMs, and the partition can be reconfigured in under a microsecond.
What it cannot isolate, however, is HBM bandwidth, since the SM subsets still share a single memory bus, 
and whichever process issues more traffic displaces the other's.

Software SM partitioning is a natural fit for the encode-side underutilization identified above.
\textsc{Encode} and \textsc{Prefill} are both compute-bound transformer forward passes, 
so co-locating them under a TPC-mask split divides compute proportionally and protects encode latency, 
while sub-microsecond reconfiguration lets the split track the encode batch size as it changes online.
The same primitive, however, also imposes two constraints that shape the rest of our design.
First, the HBM-bandwidth limitation rules out placing \textsc{Decode} on the encode GPU, 
since it streams the KV cache token by token and would saturate the shared bus regardless of the SM split. 
\textsc{Decode},  therefore,  remains on a dedicated GPU in our design.
Second, the right partition ratio depends on the current encode batch bound, which varies online,
making static split ineffective.
Specifically, the resulting split also has to absorb the local prefill load determined by the offload ratio,
which couples the SM-split decision to the dispatch and offload mechanisms developed next.

\begin{figure}[t!]
  \centering
  \includegraphics[width=1\linewidth]{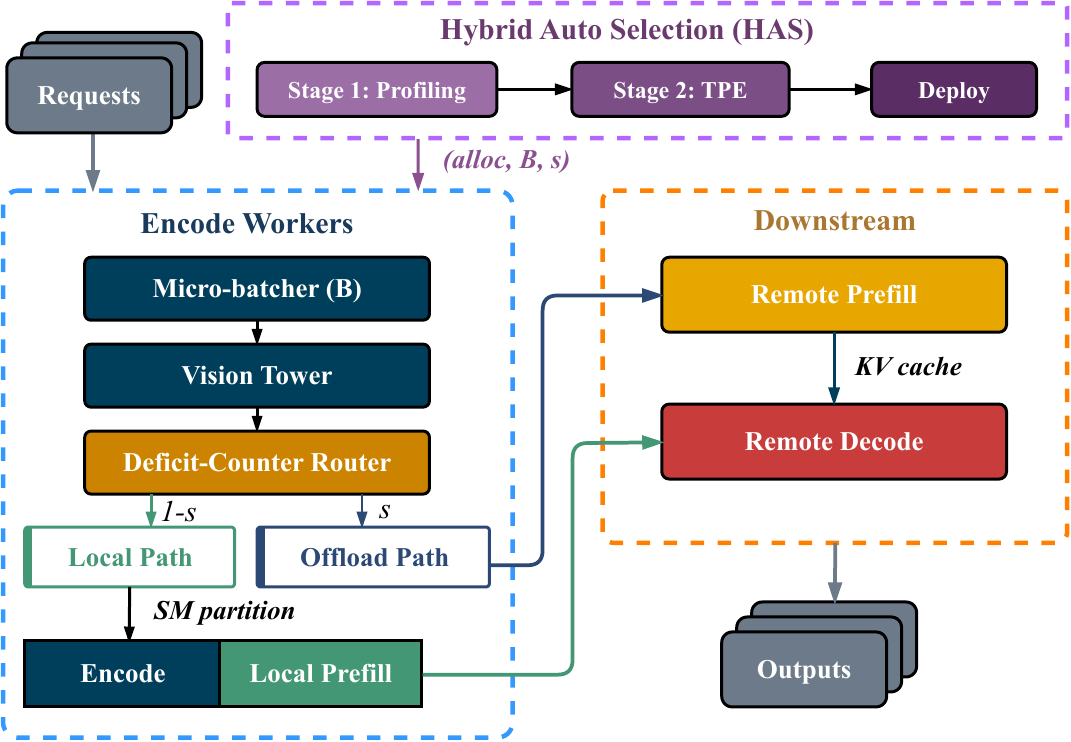}
  \caption{Overview of \projectname.}  %
  \Description{Architecture diagram of the proposed system with three grouped regions. A Hybrid Auto Selection box at the top runs Stage 1 profiling, then Stage 2 TPE search, then deployment, and emits the configuration triple of stage allocation, encode batch bound B and offload ratio s. The Encode Workers region on the left passes incoming requests through a micro-batcher governed by B, then a vision tower, then a deficit-counter router. The router splits traffic between a local path carrying fraction one minus s and an offload path carrying fraction s. The local path feeds a GPU whose SMs are split between encode and local prefill. The Downstream region on the right contains remote prefill feeding remote decode through the KV cache. Both paths converge on the outputs.}
  \label{fig:overview}
\end{figure}
\section{System Design}
\label{sec:design}

Motivated by the discussion in previous section, 
 \projectname is an encode-aware EPD serving framework organized in two co-designed layers. As shown in Figure~\ref{fig:overview}, at the top, \emph{Hybrid Auto Selection} (HAS) searches the joint
configuration $PC = (\mathrm{alloc}, B, s)$ offline,
jointly choosing the GPU allocation, encode batch bound, and prefill offload ratio.
At the bottom, three encode-aware runtime mechanisms realize the chosen
$PC$ online by controlling \emph{when} encode dispatches,
\emph{where} each request's prefill runs,
and \emph{how} the encode GPU is shared with a co-resident local prefill worker.
We describe each layer in detail in the following subsections, and
Table~\ref{tab:design-notation} lists the symbols used in the rest of this section.

\subsection{Hybrid Auto Selection}
\label{sec:design-has}

This critical layer, abbreviated as 
HAS,  automates the selection of the configuration (or $PC$) for each deployment target,
replacing the manual configuration that operators would otherwise perform.
Given the GPU budget (or count)  $G$, the model $\mathcal{M}$, the workload $\mathcal{D}$,
and the target arrival rate $\Lambda$,
HAS returns the deployment configuration that maximizes throughput:
\begin{equation}
PC^\star \;=\; \arg\max_{PC}\; \mathrm{Throughput}(PC),
\label{eq:has-objective}
\end{equation}
with ties broken by mean TTFT, then mean TPOT.
The search runs once offline, for each combination of GPU budget, model, workload, and target-rate,
and its output is then used for configuring the system at the runtime.

{Existing serving stacks automate only parts of this search:
text-only PD systems such as DistServe~\cite{zhong2024distserve}
and Splitwise~\cite{patel2024splitwise}
search stage allocation but have no \textsc{Encode} to tune,
while EPD frameworks such as Dynamo~\cite{nvidia2025dynamo} run \textsc{Encode}
as a pass-through with $B$ and $s$ fixed to 1 rather than exposed as tunable parameters.}
Searching for the configuration manually is impractical here, because the three components of $PC$ are tightly coupled:
{the SM split that protects encode latency depends on the encode batch bound $B$;
the SM partitioner enforces this split at runtime.
The  SM split, in turn,  shifts the offload ratio $s$ that keeps the pipeline balanced,
since the local prefill worker's compute capacity depends on its SM share};
and any change to $\mathrm{alloc}$ rebalances all three simultaneously.
Even on our 8-GPU testbed, the resulting space contains hundreds of valid configurations,
and the count grows exponentially with the GPU budget.
Exhaustive search is prohibitive because every candidate
requires a full deployment trial of several minutes {on our 8-GPU testbed},
adding up to hours or longer for a single workload.  
Purely analytical modeling can avoid  this cost, but it is very challenging 
to capture model and hardware characteristics fully. 

Our key insight, and the central design contribution of HAS,
is that the configuration space has an internal asymmetry
that can be exploited to combine  analytical reasoning and empirical search  effectively. 
The first-order effect of $\mathrm{alloc}$ is monotonic
and predictable from per-stage capacity alone:
if any stage is starved, the pipeline has a  bottleneck  there
regardless of how $B$ and $s$ are chosen,
so such allocations can be ruled out without ever being deployed.
The interaction between $B$ and $s$, by contrast, is non-monotonic
within the  set of feasible  allocations
and is best located by direct end-to-end measurement.
HAS therefore splits the work between two stages (Algorithm~\ref{alg:has}):
Stage~1 (lines 1--12) profiles each stage's per-worker capacity in parallel on disjoint GPU subsets,
discards allocations whose dominant stage already exceeds capacity at the target rate,
and retains the top-$K$ candidates by a bottleneck-balance score.
Stage~2 (lines 13--20) seeds a Tree-structured Parzen Estimator (TPE) 
study around each survivor's analytical optimum and
refines it through end-to-end trials, stopping once the top configurations stabilize.
This split is what lets HAS span the full coupled $(\mathrm{alloc}, B, s)$ space
in tens of trials rather than hundreds of full-system deployments,
ranking configurations by throughput, with mean TTFT and mean TPOT as tie-breakers
when configurations achieve comparable throughput.

\begin{algorithm}[t]
\small
\caption{Hybrid Auto Selection (HAS)}
\label{alg:has}
\SetAlgoNoLine
\DontPrintSemicolon
\providecommand{\algcomm}[1]{\textcolor{green!45!black}{\textnormal{#1}}}
\SetCommentSty{algcomm}
\SetKwComment{tcc}{\textcolor{green!45!black}{\#\ }}{}
\KwIn{GPU budget $G$,\; model $\mathcal{M}$,\; workload $\mathcal{D}$,\; target rate $\Lambda$}
\KwOut{Best configuration $(\mathrm{alloc}^\star,\, B^\star,\, s^\star)$}

\BlankLine
\tcc{\textbf{Stage 1:} Parallel profiling and screening}
Profile the encode saturation capacity $C_E$ and the per-worker prefill and decode capacities $C_P$ and $C_D$ on disjoint GPU subsets in parallel\;
\ForEach{\textnormal{feasible allocation} $a$ \textnormal{with} $E{+}P{+}D{=}G$}{
  {$\rho_E \leftarrow (\Lambda / n_E)\, /\, C_E$\;}
  $\rho_P \leftarrow \Lambda \,/\, (C_P\,(n_P + n_L\,\alpha))$\;
  {$\rho_D \leftarrow (\Lambda / n_D)\, /\, C_D$\;}
  $\mathrm{score}(a) \leftarrow \max(\rho_E,\, \rho_P,\, \rho_D)$\;
  {$B_{\max}(a) \leftarrow \lceil \Lambda / n_E \rceil$\;}
  {$s^*(a) \leftarrow C_{\mathrm{remote}}(a) \,/\, (C_{\mathrm{remote}}(a) + C_{\mathrm{local}}(a))$\;}
}
Discard allocations where $\mathrm{score} \ge 1$ \tcc{infeasible: overloaded stage}
Prune allocations where $\mathrm{score} > 1.2 \times \min_a(\mathrm{score})$\;
$\mathcal{A}_K \leftarrow$ top-$K$ allocations by score \tcc{$K{=}3$ in our deployment}

\BlankLine
\tcc{\textbf{Stage 2:} End-to-end refinement via TPE}
Initialize Optuna study {with per-alloc search ranges: for each $a \in \mathcal{A}_K$, $B \in [1, B_{\max}(a)]$ and $s \in [s^*(a){-}0.2,\; s^*(a){+}0.2]$}\;
\For{$t \leftarrow 1$ \KwTo $N_{\max}$}{
  $(\mathrm{alloc}, B, s) \leftarrow$ TPE-sampled configuration\;
  Deploy $(\mathrm{alloc}, B, s)$;\; run benchmark $\mathcal{D}$ at rate $\Lambda$\;
  Record throughput $y_t$ and mean (TTFT, TPOT) as tie-breakers\;
  \lIf{top-3 spread on $y_t < 1\%$ or $5$ rounds without improvement}{break}
}
\Return $\argmax$ by lexicographic order: $y_t \succ -\mathrm{TTFT}_{\mathrm{mean}} \succ -\mathrm{TPOT}_{\mathrm{mean}}$\;
\end{algorithm}

\compactparagraph{Stage~1: Profiling and Pruning.}
The role of Stage~1 is to analytically  answer the question {\em  ``is this allocation worth considering?''}. 
The enabling observation is that per-stage capacity is a property
of the combination of model, hardware, and workload, and is independent of how the GPUs are partitioned. 
Thus,
each stage can be profiled once on a small GPU subset
and the resulting capacities reused to score every candidate allocation.
Encode profiling adaptively doubles $B$ until throughput saturates,
producing a saturation capacity $C_E$ that captures the batching response. 
Prefill and Decode profiling each run a stage-targeted stress workload
on an isolated worker, yielding per-worker capacities $C_P$ and $C_D$.
The three profiles run in parallel on disjoint GPU subsets
and together cost less than a single end-to-end trial.

Given these profiles, we model each stage as an independent open-queue server
with utilization
\begin{equation}
\rho_\sigma = \frac{\Lambda\,/\,n_\sigma}{C_\sigma},\qquad \sigma\in\{E,P,D\},
\label{eq:rho_proxy}
\end{equation}
where $n_\sigma$ is the number of workers assigned to stage $\sigma$.
For prefill, $\rho_P = \Lambda / (C_P\,(n_P + n_L\,\alpha))$, where $n_L$ is the number of co-resident local prefill workers and $\alpha \in [0, 1]$ is the fraction of encode-GPU memory budget allocated to a co-resident local prefill worker, used as a memory-bound proxy for its per-worker capacity. The local term $n_L\,\alpha\,C_P$ and the remote term $n_P\,C_P$ are kept separate so that the formula remains correct under any $(n_E, n_L, n_P)$ allocation. Stage~2 absorbs any deviation between this analytical estimate and the realized local capacity.
The $1.2\times$ slack used below leaves room for Stage~2 to recover near-miss allocations.
We summarize each allocation by its bottleneck-balance score
\begin{equation}
\mathrm{score}(a) = \max(\rho_E,\;\rho_P,\;\rho_D),
\label{eq:bottleneck}
\end{equation}
and apply three filters in sequence. 
First, we 
discard allocations with $\rho_\sigma \ge 1$, since here the stage lacks the capacity to sustain the offered load. Second, we
prune those whose score exceeds $1.2\times$ the minimum
(the 20\% gap balances false pruning against wasted Stage~2 trials).  
Finally, we only retain the top-$K$ remaining allocations as $\mathcal{A}_K$ ($K{=}3$ in our deployment).

Stage~1 also seeds the parameter ranges that Stage~2 will explore, {computed per allocation $a \in \mathcal{A}_K$}:
$B$ is capped at $B_{\max}{(a)} = \lceil \Lambda / n_E \rceil$
(a worker is unlikely to fill batches much larger than its arrival rate),
and $s$ is centered at the load-balancing point
$s^*{(a)} = C_{\mathrm{remote}}{(a)} / (C_{\mathrm{remote}}{(a)} + C_{\mathrm{local}}{(a)})$
with $C_{\mathrm{remote}}(a) = n_P\,C_P$ and $C_{\mathrm{local}}(a) = n_L\,\alpha\,C_P$,
and a $\pm 0.2$ search window.

\compactparagraph{Stage~2: Bayesian Refinement.}
Stage~2 addresses the part of the problem that Stage~1  does not address. 
Specifically, as 
the $(B, s)$ surface within each surviving allocation is non-monotonic,
so we treat it as a black box and probe it with end-to-end trials.
Because the surface is mixed-type (categorical $\mathrm{alloc}$, discrete $B$, continuous $s$)
and trials are expensive, we use a single Optuna~\cite{akiba2019optuna} study
with the TPE sampler,
which natively handles such spaces and concentrates trials on regions
the surrogate model believes are promising.
Each trial deploys the full E+P+D system at rate $\Lambda$,
runs the benchmark workload,
and records throughput together with mean TTFT and mean TPOT for tie-breaking.
The search terminates when the spread among the top three configurations
falls below 1\% or five consecutive rounds show no improvement. 
The  final configuration is chosen lexicographically:
throughput first, then mean TTFT and TPOT.
Our evaluation confirms that HAS reaches near-optimal configurations substantially 
faster than uninformed search baselines as we will show later (\S\ref{sec:eval-has}).
The HAS hyperparameters
(pruning factor $1.2\times$, top-$K{=}3$, $s$ window $\pm 0.2$, 30-trial budget)
are fixed once and reused across all reported runs.

\begin{figure}[!t]
  \centering
  \includegraphics[width=.92\columnwidth]{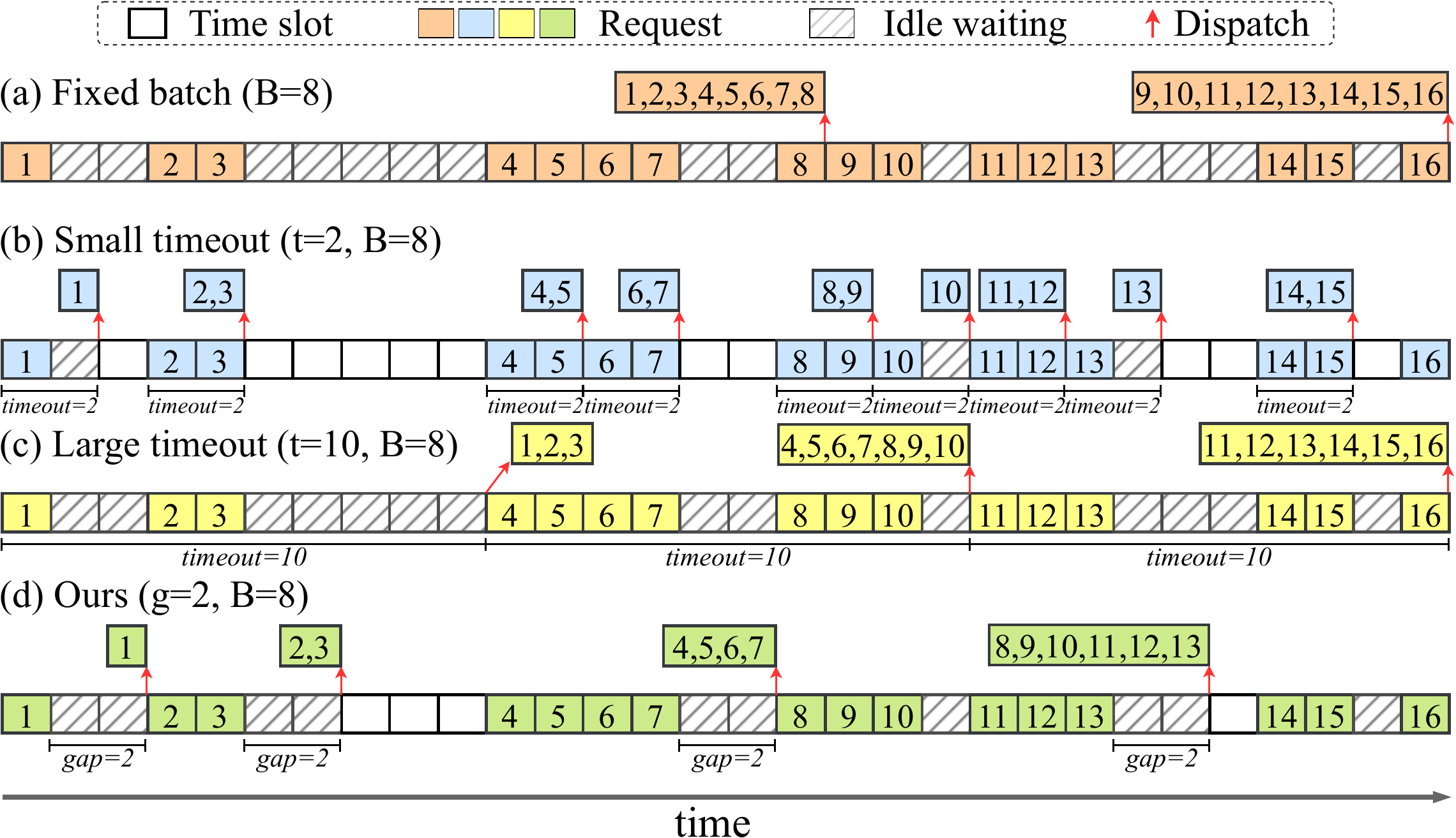}
  \caption{Four dispatch policies on the same Poisson arrival trace.
  \emph{(a)}~Fixed batch bound, no timeout.
  \emph{(b)}~Timeout too small.
  \emph{(c)}~Timeout too large.
  \emph{(d)}~Our Poisson-gap threshold $g = \kappa/\lambda$.}
  \Description{Four timelines of the same Poisson arrival trace of sixteen requests, one per dispatch policy, with hatching marking idle waiting and arrows marking dispatches. Row (a), a fixed batch bound of eight with no timeout, issues only two dispatches and leaves long idle stretches. Row (b), a timeout of two, issues nine dispatches, several carrying only one or two requests. Row (c), a timeout of ten, issues three large dispatches but makes the earliest requests wait a long time. Row (d), the proposed Poisson-gap threshold with g equal to two, issues four dispatches whose sizes track the observed arrival density, cutting idle waiting without fragmenting batches.}
  \label{fig:poisson-batching}
\end{figure}

\subsection{Load-Adaptive Micro-Batching}
\label{sec:design-encode}

HAS sets the encode batch bound $B$ offline;
the micro-batcher's runtime task is to decide
\emph{when} to dispatch a partial batch that has not yet reached $B$.
Each forward pass of a vision encoder at small $B$ is dominated by kernel-launch
and scheduling overhead, so batching is necessary for  achieving high throughput. 
At the same time,  every additional request the dispatcher waits for adds latency to the critical path.
An effective dispatch policy needs to identify this trade-off point automatically across 
different arrival rates. 

We want a threshold that tightens under high load and relaxes under low load,
without per-workload retuning.
Standard dispatch policies~\cite{crankshaw2017clipper,nvidia2025dynamo}  fall short here.
A fixed timeout grows mistuned whenever load shifts.
Queue-length triggers cannot tell a transient lull from the end of a burst,
holding batches open too long at low load.
Continuous batching~\cite{yu2022orca} sidesteps the question entirely,
since it applies only to autoregressive decoding.

\compactparagraph{Poisson-gap dispatch rule.}
Our key insight is that the dispatch decision is fundamentally a property
of the arrival process itself,
so we derive the dispatch threshold directly from that process
rather than treating it as a free parameter.
Let $\lambda = \Lambda / n_E$ be the per-encode-worker arrival rate,
and let $p \in (0, 1)$ be the target probability of filling a batch before dispatch.
Modeling arrivals as Poisson
(\S\ref{sec:eval-sensitivity} reports sensitivity under non-Poisson arrivals),
inter-arrival times are exponentially distributed:
$\Pr(X \le g) = 1 - e^{-\lambda g}$.
Starting from the first enqueued request,
let $X_1, \ldots, X_{B-1}$ denote the gaps between successive arrivals;
we require each of these gaps to be at most $g$ for the batch to fill:
\[
\Pr\bigl(\max\{X_1,\ldots,X_{B-1}\}\le g\bigr)
  = \bigl(1-e^{-\lambda g}\bigr)^{B-1}.
\]
Setting this equal to $p$ and solving:
\begin{equation}
\kappa := -\ln\!\bigl(1 - p^{\frac{1}{B-1}}\bigr),
\qquad
g := \frac{\kappa}{\lambda}.
\label{eq:poisson_gap}
\end{equation}
The result has the desired property $g \propto 1/\lambda$:
the threshold tightens automatically under high load,
when batches fill quickly. 
It also relaxes under low load,
when latency pressure is lower,
with no per-workload retuning.
We assume $\lambda$ is known from the deployment's target arrival rate;
in production where $\lambda$ drifts, $\kappa$ can be recomputed periodically
from a windowed arrival-rate estimate. 
The single tuning constant $p$ trades batch fullness against tail latency;
we use $p{=}0.6$ throughout, which lies at the knee of this trade-off,
and performance is insensitive to the choice within $[0.5, 0.7]$.

Given $g$, the batcher dispatches when any of three conditions is satisfied:
the batch reaches size $B$, the idle-gap timer (reset on every arrival) exceeds $g$,
or the batch's age from its first request exceeds a hard cap $A = 1000$\,ms.
The hard cap is not intended for the common case;
it bounds the worst-case TTFT under pathological arrival patterns
in which each request arrives just before $g$ expires
and would otherwise keep a batch open for $g (B{-}1)$ seconds.
The complete dispatch condition is
\[
\text{dispatch when: } |B_{\text{cur}}| = B
  \;\lor\; \text{idle} > g
  \;\lor\; \text{age} > A,
\]
where $B_{\text{cur}}$ is the set of requests currently in the micro-batch.
Figure~\ref{fig:poisson-batching} shows how the rule detects burst boundaries
and groups arrivals into variable-sized batches without manual tuning;
the batcher adds $O(1)$ work per arrival and incurs sub-microsecond timer overhead.

\subsection{Rate-Controlled Partial Offload}
\label{sec:design-offload}

Once the micro-batcher has decided \emph{when} an encode batch dispatches,
the next question is \emph{where} each request's prefill should run.
Partial offload realizes this routing decision through encode's spatial tuning parameter $s \in [0, 1]$,
the target fraction of requests sent to a remote \textsc{Prefill} worker
rather than to a co-resident local prefill worker on the encode GPU.
We co-locate prefill, rather than decode, with encode for the reason established in \S\ref{sec:bg-mps};
remote \textsc{Prefill} and all \textsc{Decode} workers therefore run on dedicated GPUs.

\begin{figure}[!t]
  \centering
  \includegraphics[width=1\linewidth]{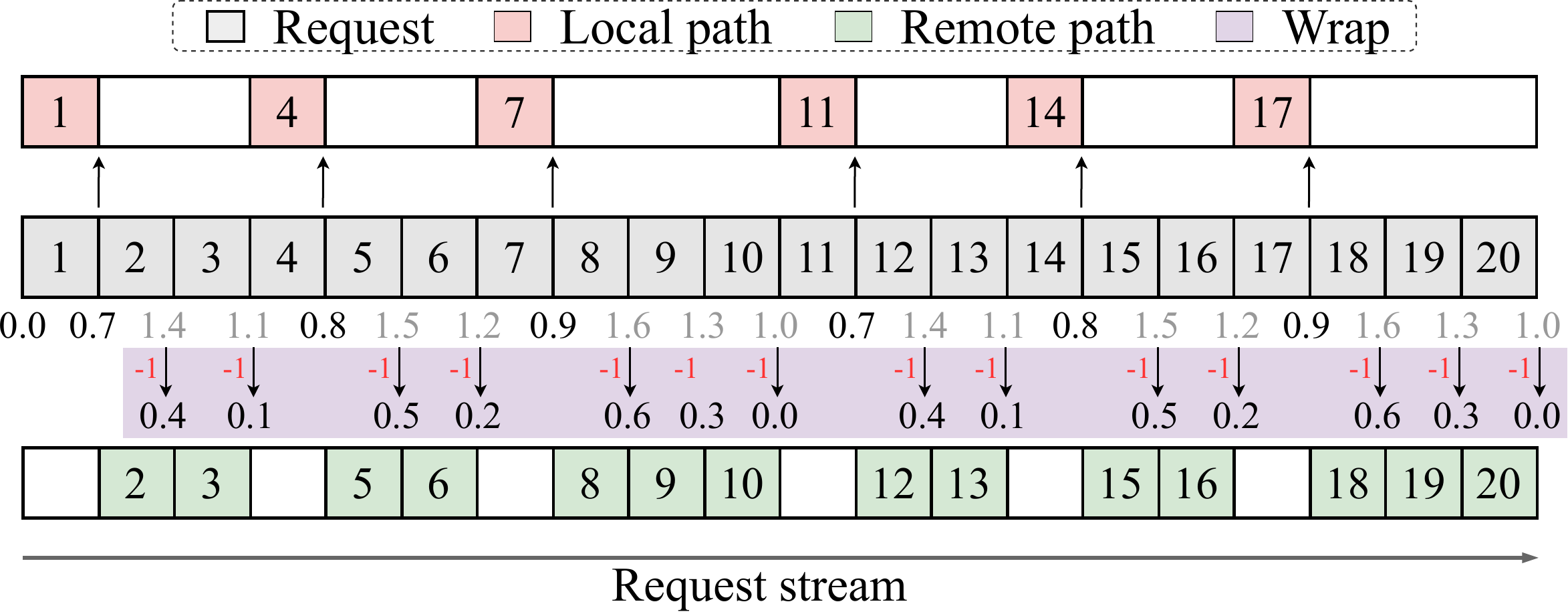}
  \caption{Deficit-counter routing with offload ratio $s{=}0.7$ over twenty requests.
  On each arrival the router increments $d$ by $s$ (value above the request).
  When $d{\ge}1$ (purple band), the request goes to the remote path
  and $d$ is decremented by~1 (value below the band);
  otherwise the request stays on the local path and $d$ carries over to the next arrival.}
  \Description{Diagram of deficit-counter routing at offload ratio 0.7 over a stream of twenty numbered requests. The middle row lists requests in arrival order, with the accumulator value after adding 0.7 printed beneath each one. A purple wrap band marks every arrival whose accumulator reaches one and subtracts one from it. The top row highlights requests 1, 4, 7, 11, 14 and 17, which stay on the local path. The bottom row highlights the remaining requests, which take the remote path, yielding roughly seven remote dispatches for every three local ones.}
  \label{fig:deficit-counter}
\end{figure}

The router's role is to make $s$ meaningful at runtime.
If the actual routing pattern drifts from the configured ratio,
or if requests reach one path in bursts,
the latency observed by HAS reflects routing noise rather than the intended setting.
Two natural baselines both fail this test:
randomized routing (each request goes remote with probability $s$)
is noisy at short horizons and can produce long runs of consecutive remote decisions
even when $s$ is moderate;
windowed routing (send exactly $\lfloor sW \rfloor$ of every $W$ requests remote)
trades randomness for a new tuning parameter $W$
whose appropriate value is workload-dependent.
We therefore adopt a parameter-free scheme whose worst-case routing deviation is bounded analytically.

\compactparagraph{Deficit-controlled routing.}
Our solution is a counter-based scheme inspired by Deficit Round Robin~\cite{shreedhar1996drr}
that converges exactly to $s$ in the long run
and bounds the worst-case run length per direction,
without introducing any new tuning parameter.
Where DRR maintains per-flow deficit counters to schedule packets among competing queues,
we collapse the same idea into a single scalar that splits a request stream
into a continuous-valued ratio between two paths.
The resulting bounded-spacing property
is what makes the offload ratio $s$ chosen by HAS observable as latency at runtime,
the property a black-box search like Stage~2 needs.
The router maintains a real-valued accumulator $d$, initialized to $0$;
for each request it adds $s$ to $d$,
sends the request remotely if $d \ge 1$ (and decrements $d$ by $1$),
and locally otherwise (Figure~\ref{fig:deficit-counter}).
Two properties follow.
First, after $n$ requests the cumulative remote count is $\lfloor ns + d_0 \rfloor$
for some $d_0 \in [0, 1)$,
so it stays within $\pm 1$ of $ns$ for all $n$ and converges exactly to fraction $s$.
Second, the counter never produces more than $\lceil 1/(1{-}s) \rceil$ consecutive remote decisions
or $\lceil 1/s \rceil$ consecutive local decisions,
which keeps the downstream queues smoother than randomized routing.
The router adds $O(1)$ work per request.

\compactparagraph{Envelope-based embedding transfer.}
After encode, each request's vision embedding has to be delivered to its assigned prefill worker.
Sending one embedding at a time would pay the transfer setup cost on every request.
Instead, we group all embeddings for the same prefill worker into one contiguous tensor
and ship them as a single NIXL~\cite{nixl2025} transfer per destination,
so a micro-batch of $N$ requests issues at most one transfer per prefill worker, 
rather than $N$ separate ones.
A small concurrency cap on in-flight sends provides backpressure
and bounds memory use without stalling the pipeline.

\subsection{Dynamic SM Partitioning}
\label{sec:design-sm}

\begin{figure}[!t]
  \centering
  \includegraphics[width=0.65\linewidth]{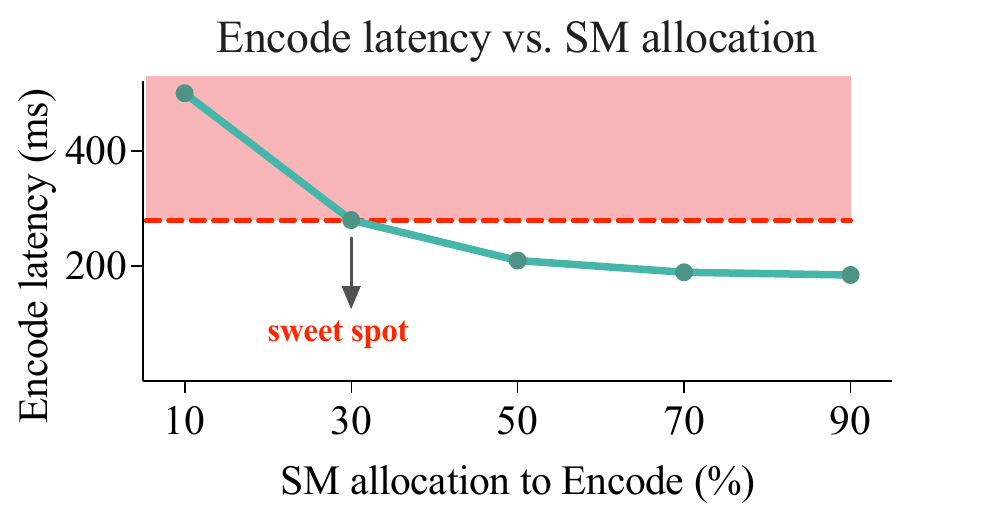}
  \vspace{-0.5em}
  \caption{Encode latency versus SM allocation on \textsc{LLaVA-v1.6-34B} at $B{=}4$.
  The dashed line marks the 20\% latency threshold
  defining the minimum allocation preserving encode QoS.}
  \Description{Line chart of encode latency in milliseconds against the share of streaming multiprocessors allocated to encode, measured on LLaVA-v1.6-34B at batch size four. Latency drops steeply from about 500 milliseconds at 10 percent to about 280 milliseconds at 30 percent, then flattens to roughly 210, 190 and 185 milliseconds at 50, 70 and 90 percent. A red dashed line near 280 milliseconds marks the 20 percent latency threshold, and the shaded band above it marks allocations that violate encode quality of service. An arrow labels 30 percent as the sweet spot.}
  \label{fig:sm-profile}
\end{figure}

\begin{figure}[!t]
  \centering
  \includegraphics[width=.92\linewidth]{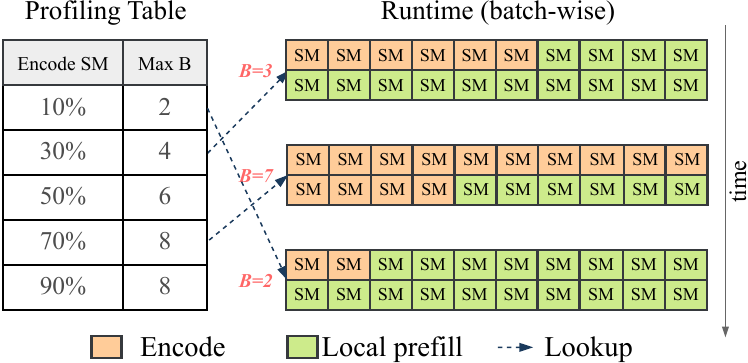}
  \caption{Dynamic SM partitioning.
  \emph{Left}: the offline table maps each SM level to the largest batch size.
  \emph{Right}: at runtime, the split between encode (orange) and local prefill (green)
  shifts with batch size while staying on disjoint SM subsets.}
  \Description{Two-part diagram of dynamic SM partitioning. The profiling table on the left maps each encode SM level of 10, 30, 50, 70 and 90 percent to the largest batch size it sustains, namely 2, 4, 6, 8 and 8. The runtime view on the right shows three snapshots of the SM array at successive times, each split between orange encode blocks and green local-prefill blocks. Dashed lookup arrows connect batch sizes of 3, 7 and 2 to their table rows, and the orange share of the array grows and shrinks as the batch size changes.}
  \label{fig:sm-runtime}
\end{figure}

Once the router has decided \emph{where} each request's prefill runs,
the encode GPU may host both an encode worker and a co-resident local prefill worker. 
We still need to decide \emph{how} the GPU is split between them.
This split involves the following tradeoff: 
encode is on the critical path and should not be slowed below its latency target,
while any compute encode does not use is wasted unless given to local prefill.
This balance is workload-dependent,
so the split has to be set per micro-batch rather than statically.

Our key insight is that the encode latency-versus-SM curve saturates:
encode latency drops sharply as it receives more SMs,
then flattens once it has enough to meet its target (Figure~\ref{fig:sm-profile}).
This means there is a well-defined  {\em smallest safe} SM share for each batch size,
and any SM above that point can be released to local prefill at no cost to encode quality of service (QoS).
We therefore re-split the SMs at every micro-batch:
look up the smallest safe share for the current batch,
give it to encode, and hand the remainder to local prefill.
The split adapts to batch size with no new tuning parameter.

We materialize the  {\em smallest safe share}  as a lookup table indexed by {SM-allocation level},
built once at startup.
Building it amounts to profiling encode latency
at a small set of SM-allocation levels {(5 in our deployment)} across a range of batch sizes,
and recording, {for each SM level, the largest batch size}
whose latency stays close to the exclusive-access baseline.
The cost is modest: each profiling point is a single encode-only forward pass,
so the full sweep runs in minutes on one GPU and produces only a handful of table entries.
The table depends only on the encode model, GPU, and workload,
so it is reused across deployment rates and offload ratios,
and its storage and lookup cost at runtime are negligible.

At runtime, the encode worker {selects the smallest SM share whose supported batch size covers the current micro-batch size},
and the local prefill worker takes the complementary fraction
(e.g., 70\% for encode and 30\% for prefill).
Both processes install the split through TPC masks via \texttt{libsmctrl}~\cite{libsmctrl},
which reconfigures in sub-microsecond and requires no kernel restart.
Small batches therefore release more SMs to local prefill,
while large batches reclaim them for encode (Figure~\ref{fig:sm-runtime}).
On workloads where HAS selects $s{=}1$ (e.g., video in our evaluation),
local prefill is not co-located and dynamic SM partitioning collapses to a
single-tenant encode allocation, so it incurs no overhead when its benefit is small.

\begin{table}[t]
  \centering
  \caption{Evaluated MLLM architectures.}
  \vspace{-0.5em}
  \label{tab:models}
  \small
  \resizebox{\columnwidth}{!}{%
  \begin{tabular}{l l l l r}
    \toprule
    Model & Modal. & Encoder & LLM backbone & Params \\
    \midrule
    LLaVA-v1.6-34B~\cite{liu2023llava} & Image & CLIP ViT-L~\cite{radford2021clip} & Yi-34B~\cite{ai2024yi} & 34B \\
    Qwen2.5-VL-32B~\cite{bai2025qwen25vl} & Video & ViT (dyn.) & Qwen2.5-32B & 32B \\
    Ultravox-v0.6-27B~\cite{fixie2024ultravox} & Audio & Whisper~\cite{radford2023whisper} & Gemma-2-27B~\cite{team2024gemma} & 27B \\
    \bottomrule
  \end{tabular}%
  }
\end{table}

\begin{table}[t]
  \centering
  \caption{SLO thresholds at the P99 level (TTFT in s / TPOT in ms) across five tiers.}
  \vspace{-0.5em}
  \label{tab:slo}
  \small
  \resizebox{\columnwidth}{!}{%
  \begin{tabular}{l c c c c c}
    \toprule
    Model & \makecell{S1\\(Strict)} & \makecell{S2\\(Tight)} & \makecell{S3\\(Moderate)} & \makecell{S4\\(Standard)} & \makecell{S5\\(Relaxed)} \\
    \midrule
    LLaVA & 5/70 & 40/80 & 50/120 & 90/120 & 140/120 \\
    Qwen & 5/60 & 10/70 & 40/110 & 80/110 & 110/110 \\
    Ultravox & 2/50 & 3/60 & 5/80 & 10/100 & 30/130 \\
    \bottomrule
  \end{tabular}%
  }
\end{table}

\section{Evaluation}
\label{sec:evaluation}

\subsection{Experimental Setup}
\label{sec:eval-setup}

\compactparagraph{Models and workloads.}
We evaluate three MLLM architectures spanning image, video, and audio,
with different encoders and LLM backbones (Table~\ref{tab:models}).
For image and video, we use the standard random multimodal workload generator
from the vLLM benchmarking suite~\cite{kwon2023vllm},
which produces synthetic inputs at realistic shapes
paired with random text prompts of typical length.
For audio, we draw real speech samples from LibriSpeech~\cite{panayotov2015librispeech}
with random text prompts.
This mix follows established serving-benchmark practice~\cite{zhong2024distserve,patel2024splitwise,hydrainfer2025}
and isolates serving-system performance from dataset-specific content effects.

\compactparagraph{Testbed and baselines.}
\cradd{Our primary testbed is a server}
with two AMD EPYC~9254 CPUs (48 cores, 96 threads),
1\,TB DRAM, and eight NVIDIA RTX~6000~Ada GPUs (48\,GB each, PCIe, CUDA~13.0),
and Ubuntu~22.04~LTS.
Each encode worker occupies one GPU at TP${=}1$,
while \textsc{Prefill} and \textsc{Decode} workers span 2 or 4 GPUs (TP${=}2$ or TP${=}4$),
since the 27B--34B backbones require TP${\ge}2$ to fit on a single 48\,GB GPU.
Together with the choice of encode batch bound $B$ and prefill offload ratio $s$,
the resulting joint configuration space contains hundreds of valid points even at 8 GPUs. 
Both Dynamo's manual sweep and HAS's automated search draw from the same space.
This setup captures the encode-prefill-decode imbalance our design targets. Multi-node scaling and cross-rack KV transfer will be topics for future efforts.
When local prefill is enabled, it shares the encode GPUs through SM partitioning
following the dynamic split rule of \S\ref{sec:design-sm}.
We compare four system configurations under the same 8-GPU budget:
\textit{vLLM}~(v0.14.1) and \textit{Dynamo}~(v0.9.0);
\textit{Ours ($s{=}1$)}, {our system with $s$ fixed to 1 (no local prefill)}; and
\textit{Ours (HAS-tuned)}, {our system with HAS-tuned $s$ (local prefill enabled with dynamic SM partitioning when $s{<}1$)}.
All baselines use their best-performing configurations on this testbed.
\cradd{We further evaluate \projectname against two EPD-serving systems,
EPDServe~\cite{epdserve2025} and HydraInfer~\cite{hydrainfer2025},
on two eight-GPU platforms: RTX~6000~Ada and A100~SXM4
(Table~\ref{tab:platform-baselines}).}

\compactparagraph{Metrics and SLOs.}
Following prior LLM-serving work~\cite{zhong2024distserve,sola2025},
we report goodput (Eq.~\ref{eq:goodput}), throughput,
{mean and} P99 time-to-first-token (TTFT), and {mean and} P99 time-per-output-token (TPOT).
Five SLO tiers per (model, workload) pair (Table~\ref{tab:slo})
range from strict to relaxed,
and goodput is computed independently at each tier.
Following standard serving-benchmark practice~\cite{zhong2024distserve,patel2024splitwise,hydrainfer2025},
each run uses Poisson arrivals at a controlled rate\cradd{.
For the main end-to-end comparison, we vary the arrival rate}
from 1 to 5\,req/s to cover under-, near-, and over-saturated regimes
(the saturation point for our three models lies in the 2--4\,req/s range).
Each run processes 500 requests;
reported numbers are medians of three independent runs.
For Dynamo, ``best manually tuned'' means sweeping the feasible allocations
and selecting the highest-goodput one at the target rate;
HAS searches the same allocation space jointly with $(B, s)$,
so the comparison differs in search mechanism rather than in hardware budget.

\subsection{End-to-End Comparison}
\label{sec:eval-e2e}

\begin{figure}[t!]
  \centering
  \includegraphics[width=\linewidth]{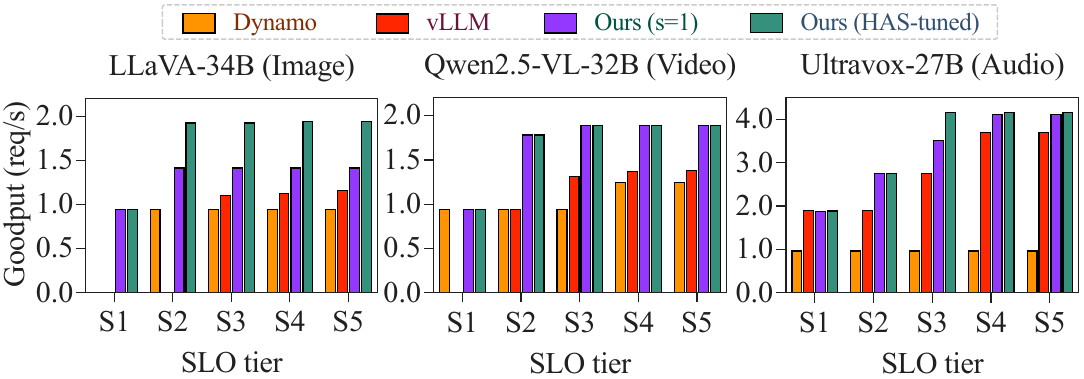}
  \caption{End-to-end goodput across five SLO tiers for image, video, and audio serving.
  The strictest tier (S1) characterizes the SLO frontier rather than an attainable operating point at the rates we sweep.}
  \Description{Grouped bar chart of end-to-end goodput in requests per second across five SLO tiers labeled S1 through S5, with one panel per model: LLaVA-34B on image, Qwen2.5-VL-32B on video and Ultravox-27B on audio. Each tier shows four bars for Dynamo, vLLM, the proposed system at s equal to one, and the proposed system with HAS tuning. Both of the proposed configurations exceed both baselines in every tier where any system attains goodput, with the widest margin on audio, where goodput approaches 4 requests per second against roughly 1 for Dynamo. The strictest tier S1 is unattainable for most systems at the rates swept.}
  \label{fig:goodput-bars}
\end{figure}

\begin{figure}[t!]
  \centering
  \includegraphics[width=\linewidth]{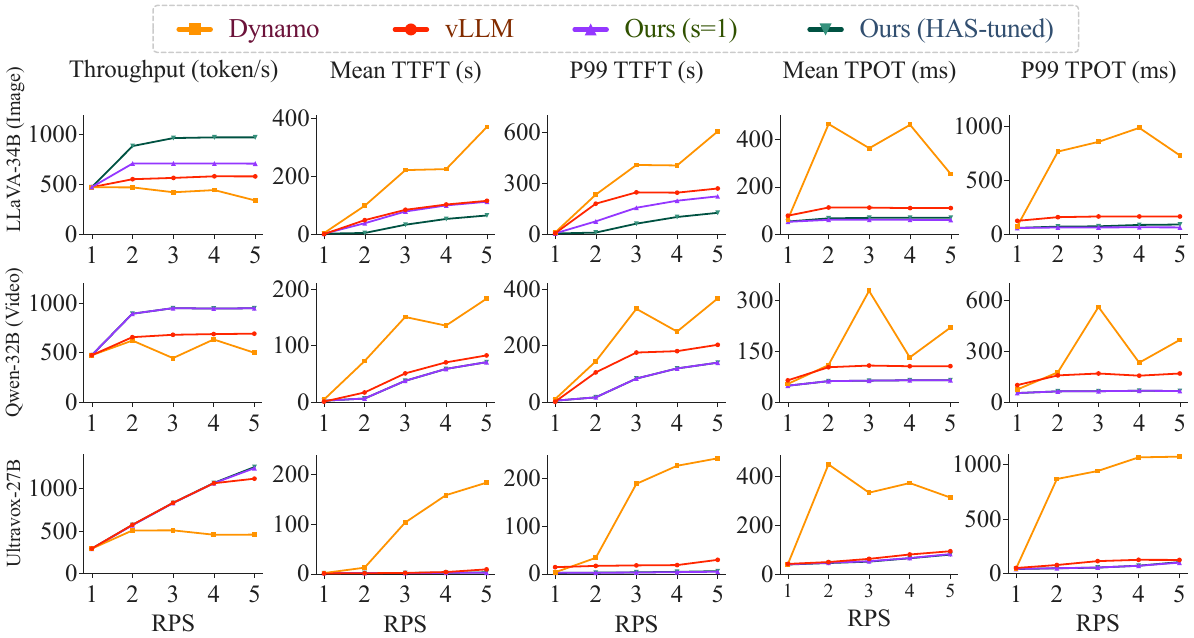}
  \caption{Throughput and latency comparison across different request rates for various modalities.
  Dynamo's TTFT diverges past saturation because its per-request encode ($B{=}1$) cannot match the offered rate;
  the encode queue grows monotonically within the run and every request inherits the accumulated wait.}
  \Description{Grid of fifteen line charts arranged in three rows, one per model, namely LLaVA-34B on image, Qwen-32B on video and Ultravox-27B on audio, and five columns showing throughput in tokens per second, mean and P99 time to first token in seconds, and mean and P99 time per output token in milliseconds. Every chart plots request rate from 1 to 5 requests per second on the horizontal axis, with four series for Dynamo, vLLM, the proposed system at s equal to one, and the proposed system with HAS tuning. The two proposed configurations hold the highest throughput and the flattest latency curves across all three modalities, while Dynamo's latency curves climb steeply once the offered rate exceeds what its per-request encode can sustain.}
  \label{fig:latency-grid}
\end{figure}

We first compare against baselines across three models and five SLO tiers.
Figure~\ref{fig:goodput-bars} reports goodput across SLO tiers,
and Figure~\ref{fig:latency-grid} reports throughput and latency under various request rates.

\compactparagraph{End-to-end results.}
\cradd{As shown in Figure~\ref{fig:goodput-bars}, 
\projectname achieves higher goodput
than vLLM and Dynamo across the three evaluated modalities,}
reaching up to $4.3\times$ higher goodput than Dynamo and $1.7\times$ higher than vLLM at the moderate tier.
The gap to Dynamo is much larger than the gap to vLLM
because Dynamo coordinates the three stages at request granularity
and does not balance encode-side compute against the downstream prefill and decode demand. 
This leaves the encode GPU as the dominant bottleneck under load. 
vLLM avoids this specific failure mode through continuous batching at the LLM,
so the headroom we improve over it is smaller.
The relative performance between Dynamo and vLLM also depends on the modality:
Dynamo {beats vLLM on video only at the strictest tier}, where the multi-frame encoder dominates the pipeline
and disaggregating it pays off. 
Dynamo, however, is slower than vLLM on audio, where the encoder is light
and aggregated serving handles the decode-heavy workload more efficiently.
EPD disaggregation alone therefore does not consistently beat aggregated serving;
our encode-aware design closes this gap and delivers consistent gains over the baselines.

Most of the gain comes from adaptive batching and stage allocation
(Figure~\ref{fig:goodput-bars}),
with local prefill adding a smaller, modality-dependent gain we examine below.
The breakdown by modality is as follows.
On image, adaptive batching plus selective local prefill (with dynamic SM partitioning)  
are most effective. 
Video is dominated by stage allocation and adaptive batching, since HAS selects $s{=}1$ and skips local prefill entirely.
For audio, the dominant factor is selective local prefill on top of the light Whisper encoder.

At the moderate tier, \projectname raises image goodput from 1.11~req/s (vLLM)
and 0.95~req/s (Dynamo) to 1.93~req/s,
reaches 1.90~req/s on video against 1.32 / 0.95,
and 4.17~req/s on audio against 2.77 / 0.98.
Tail latencies drop sharply at the same time:
at $\Lambda{=}2$\,req/s, P99 TTFT drops from Dynamo's 233.8\,s to 10.8\,s on image
and from 144.8\,s to 18.0\,s on video.
Dynamo's  performance is so slow  because its encode stage runs over saturation at this rate,
so the encode queue grows unboundedly within the run
and every request inherits the accumulated wait;
\projectname's encode-aware dispatch keeps the encoder at line rate, the queue never builds up,
and the tail collapses to roughly the prefill+decode time alone.
P99 TPOT drops in step (764$\to$72\,ms on image, 869$\to$49\,ms on audio)
for the same reason: once encode keeps up with downstream demand,
decode batches no longer stall mid-run, and TPOT returns to its native rate.
The composition of these gains is itself modality-dependent.
For image and video, encode is the upstream bottleneck,
so adaptive batching alone is enough to shorten the encode--prefill tail and bring the  TTFT value down.
For audio, the encoder is light and the workload is decode-heavy,
so the dominant factor is selective local prefill,
which relieves downstream queueing and surfaces as a gain in TPOT and throughput.

\begin{figure}[t!]
  \centering
  \includegraphics[width=0.92\linewidth]{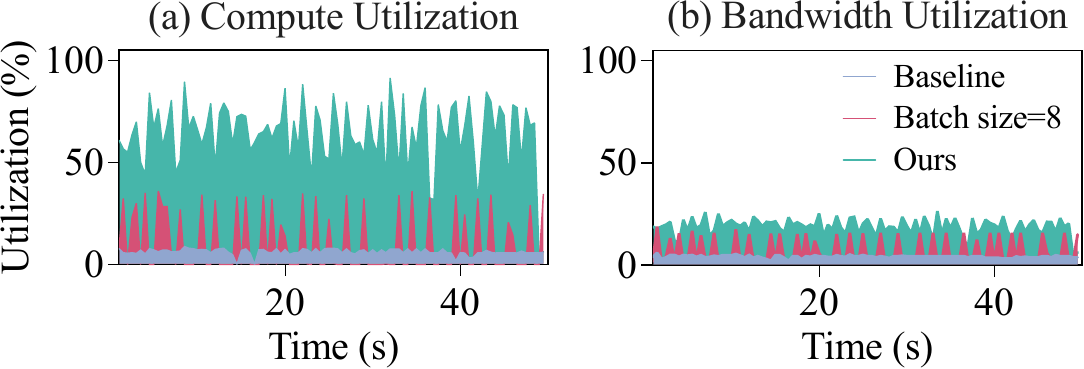}
  \caption{GPU utilization comparison against baseline.}
  \Description{Two time-series plots over a 50-second window on a vertical axis running from 0 to 100 percent, extending the earlier baseline comparison with the proposed system. In panel (a), compute utilization for the proposed system oscillates between roughly 40 and 90 percent, the batch-size-8 curve spikes to about 20 percent, and the baseline stays near zero. In panel (b), bandwidth utilization for the proposed system holds around 15 to 20 percent while both baselines remain lower. The proposed system keeps the encode GPU busy for most of the run.}
  \label{fig:encode-utilization-ours}
\end{figure}

Figure~\ref{fig:encode-utilization-ours} further shows where these gains come from.
Dynamo's request-granularity coordination keeps the encode GPU below $10\%$ utilization,
and naive fixed batching raises it only intermittently.
\projectname fills the remaining idle gaps with co-located local prefill,
raising sustained utilization to roughly $80\%$.
This means the remote \textsc{Prefill} queue empties faster,
which is the main reason TTFT drops so sharply in Figure~\ref{fig:latency-grid}.

\begin{table}[t!]
  \centering
  \crcolor
  \ifcrmarked
    \captionsetup{labelfont={bf,color=red},textfont={bf,color=red}}
  \fi
  \caption{Throughput and latency comparison across two eight-GPU platforms
  on LLaVA-v1.6-34B at $\Lambda{=}\crplatformrate$\,req/s, using the
  best-performing configuration of each system.}
  \label{tab:platform-baselines}
  \small
  \setlength{\tabcolsep}{3.3pt}
  \begin{tabular*}{\columnwidth}{@{\extracolsep{\fill}}lrrrrr@{}}
    \toprule
    System & Tput & \multicolumn{2}{c}{TTFT (s)} & \multicolumn{2}{c}{TPOT (ms)} \\
           & (req/s) & Mean & P99 & Mean & P99 \\
    \midrule
    \multicolumn{6}{@{}l}{\textit{RTX~6000~Ada 48\,GB (PCIe)}} \\
    EPDServe   & 0.46 & 448 & 929 & 119 & 137 \\
    HydraInfer & 0.58 & 315 & 664 & 402 & 477 \\
    \projectnamenott & \textbf{1.95} & \textbf{66} & \textbf{127} & \textbf{71} & \textbf{91} \\
    \midrule
    \multicolumn{6}{@{}l}{\textit{A100 SXM4 80\,GB (NVLink)}} \\
    EPDServe   & 1.01 & 142 & 304 & 249 & 330 \\
    HydraInfer & 1.51 & 118 & 214 & 128 & 233 \\
    \projectnamenott & \textbf{4.91} & \textbf{16} & \textbf{29} & \textbf{91} & \textbf{105} \\
    \bottomrule
  \end{tabular*}
\end{table}

\compactparagraph{Modality-dependent behavior.}
The two ``Ours'' bars in Figure~\ref{fig:goodput-bars} split the gain into two parts.
Dynamo vs.\ Ours ($s{=}1$) isolates adaptive micro-batching and HAS-tuned stage allocation,
which accounts for a substantial portion of the improvement across modalities;
Ours ($s{=}1$) vs.\ Ours (HAS-tuned) isolates the additional effect of local prefill.
How much local prefill helps depends on the modality, and HAS chooses automatically.
On image, the CLIP encoder is heavy enough to benefit from disaggregation
yet light enough to leave SM headroom for local prefill.
Therefore,  HAS picks $s{<}1$ and goodput rises from 1.42 to 1.93~req/s.
On audio, the Whisper encoder leaves substantial headroom and HAS again picks $s{<}1$.
On video, the multi-frame encoder uses more than $90\%$ of the SMs during batched forwards.
Since SM partitioning does not isolate HBM bandwidth (\S\ref{sec:bg-mps}),
co-locating prefill would slow encode rather than help,
and HAS correctly picks $s{=}1${; Ours (HAS-tuned) therefore coincides with Ours ($s{=}1$) on video, and the two ``Ours'' bars and lines overlap in Figures~\ref{fig:goodput-bars} and~\ref{fig:latency-grid}}.

\begin{crrevision}
\compactparagraph{Cross-platform comparison.}
Beyond the vLLM and Dynamo results above, 
we further compare \projectname with
EPD-serving systems EPDServe~\cite{epdserve2025} and
HydraInfer~\cite{hydrainfer2025} on LLaVA-v1.6-34B
(Table~\ref{tab:platform-baselines}).
We run the same image workload at $\Lambda{=}\crplatformrate$\,req/s
on the RTX~6000~Ada testbed and on eight A100 SXM4 GPUs
(80\,GB HBM2e each, NVLink).
We use the latest public revisions available at evaluation time: EPDServe
commit \texttt{9b111f7} and HydraInfer v0.1.0 (commit \texttt{009257f}).
For each system, we select its best-performing configuration separately
on each platform rather than carrying one configuration across GPU types.
On RTX~6000~Ada, \projectname reaches 1.95\,req/s,
$4.24\times$ EPDServe and $3.36\times$ HydraInfer.
On A100, it reaches 4.91\,req/s, maintaining throughput advantages of
$4.86\times$ and $3.25\times$, respectively.
The latency ordering is also consistent: \projectname has lower mean
and P99 TTFT and TPOT than both baselines on both platforms.
For example, its P99 TTFT is 127\,s on RTX~6000~Ada, compared with
929\,s for EPDServe and 664\,s for HydraInfer; on A100, it falls to
29\,s, compared with 304\,s and 214\,s, respectively.
Thus, the advantage is preserved on both the PCIe-based RTX~6000~Ada
platform and the NVLink-connected A100 platform.
\par
\end{crrevision}

\begin{figure}[t!]
  \centering
  \includegraphics[width=\linewidth]{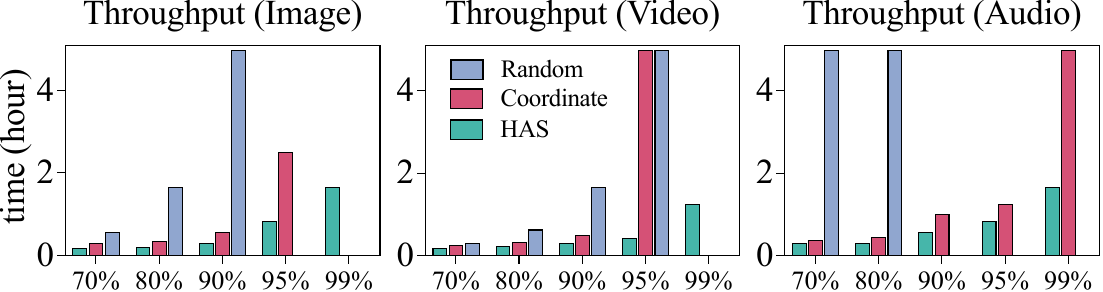}
  \caption{{Wall-clock time HAS, Random, and Coordinate each take to reach a given percentile of the best-observed throughput.} The per-modality reference is the maximum across all 3 methods $\times$ 30 trials; missing bars indicate the percentile was not reached within the 30-trial budget.}
  \Description{Grouped bar chart of the wall-clock hours each search method needs to reach a given percentile of the best observed throughput, with one panel per modality for image, video and audio. The horizontal axis lists the 70th, 80th, 90th, 95th and 99th percentiles and the vertical axis runs from 0 to 5 hours. Three series are plotted for Random, Coordinate and HAS. HAS reaches the higher percentiles in less wall-clock time than both baselines in most cells, and it reaches percentiles that the baselines never attain. Missing bars mark percentiles a method failed to reach within its budget of 30 trials.}
  \label{fig:has-time}
\end{figure}

\subsection{HAS Efficiency}
\label{sec:eval-has}

We evaluate HAS as the configuration layer that selects
$(\text{allocation},\,B,\,s)$ for each deployment.
This search is non-trivial because the joint $(\mathrm{alloc}, B, s)$ space
contains hundreds of configurations even at 8 GPUs,
its dimensions interact non-monotonically,
and the best operating point varies with modality and load.
We therefore compare HAS against two search baselines under the same budget of
30~trials. Each trial deploys one configuration and runs 100 requests at
$\Lambda{=}5$\,req/s; we report results on the image (\textsc{LLaVA-34B}), video (\textsc{Qwen-32B}), and audio (\textsc{Ultravox-27B}) workloads.
All methods search the same joint space of feasible allocations, micro-batch
sizes, and offload ratios: \emph{Random} samples uniformly from this space,
whereas \emph{Coordinate} performs coordinate descent over the three dimensions.
For all three methods, the selected configuration is the one with the highest
{throughput observed within the 30-trial budget, with mean TTFT and mean TPOT as tie-breakers},
and the reported throughput and latency metrics correspond to that configuration.
Figure~\ref{fig:has-time} reports the wall-clock search cost of each method.

\compactparagraph{Stage 1 pruning.}
Run at the same target rate as Stage 2 ($\Lambda{=}5$\,req/s, chosen above saturation
to expose method differences),
Stage 1's bottleneck-balance score (Eq.~\ref{eq:bottleneck})
prunes the allocation dimension of the search space,
keeping 3 of the 7 feasible allocations under the $1.2\times$ threshold.
The four pruned allocations all have a bottleneck stage at least 20\% more loaded
than the best-balanced candidate at this rate,
so that stage would dominate end-to-end latency regardless of how $B$ and $s$ are tuned.
This $7{\to}3$ reduction lets Stage 2 concentrate its 30-trial budget
on balanced candidates, improving per-trial reliability.

\compactparagraph{{Search quality.}}
{Across the three modalities, HAS consistently finds the highest-throughput configuration within the 30-trial budget.
The benefit is largest on audio, where the light Whisper encoder leaves substantial encode-GPU headroom; HAS's directed search exploits this headroom more reliably than uninformed baselines;
moderate on image, where the CLIP encoder is heavy enough to benefit from disaggregation but still leaves enough headroom for local prefill;
and smallest on video, where the multi-frame encoder leaves little room for local prefill and narrows the useful part of the configuration space.}

\compactparagraph{{Search efficiency.}}
{Figure~\ref{fig:has-time} reports the average wall-clock time each method takes to find a configuration reaching a given percentile of the best-observed throughput across the three modalities.
HAS reaches the 99th percentile in 1.7~h on image, 1.3~h on video, and 1.7~h on audio, while Random fails to reach the 99th percentile on any modality within its 30-trial budget, and Coordinate fails on image and video.
Even at lower percentiles HAS is consistently faster: it reaches the 90th percentile in 0.3~h on image (vs.\ 5.0~h for Random and 0.6~h for Coordinate) and 0.6~h on audio (where Random does not reach this percentile within its budget).
These differences also reveal each baseline's failure mode: Random's slow rise to high percentiles suggests it spends much of its budget in poor regions of the search space, while Coordinate descends quickly at low percentiles but stalls at high ones, consistent with descent into a local optimum.
HAS avoids both behaviors by pruning bottlenecked allocations before search, so its 30-trial budget concentrates on balanced candidates.}

\begin{table}[t!]
  \centering
  \caption{SM partitioning ablation on \textsc{Ultravox-27B} with local prefill enabled. All three configurations share allocation 2E(TP1)+1P(TP2)+1D(TP4), $B{=}10$, $s{=}0.55$, and $\Lambda{=}10$\,req/s.}
  \label{tab:sm-ablation}
  \small
  \setlength{\tabcolsep}{3.8pt}
  \begin{tabular}{l r r r}
    \toprule
    SM config & Tput (r/s) & P99 TTFT (s) & P99 TPOT (ms) \\
    \midrule
    No partition (time-slicing) & 4.78 & 43.7 & 236 \\
    Static 50/50                 & 5.07 & 40.7 & 203 \\
    Dynamic (ours)               & \textbf{5.21} & \textbf{36.7} & \textbf{167} \\
    \bottomrule
  \end{tabular}
\end{table}

\subsection{SM Partitioning Ablation}
\label{sec:eval-sm-ablation}

We isolate the contribution of dynamic SM partitioning by comparing three configurations on \textsc{Ultravox-27B} with local prefill enabled:
(i) no SM partitioning (default CUDA time-slicing between encode and local prefill),
(ii) static 50/50 SM split via \texttt{libsmctrl}, and
(iii) our dynamic five-level scheme.

Table~\ref{tab:sm-ablation} reports the results at $\Lambda{=}10$\,req/s{, a high rate that pushes all three configurations past their throughput limit so the throughput column shows what each can actually sustain}.
Figure~\ref{fig:sm-profile} explains why dynamic partitioning is effective:
encode latency decreases sharply as the encode worker receives more SMs,
but the benefit quickly saturates past the latency knee.
This diminishing-return profile implies that a fixed 50/50 split is rarely optimal;
once encode receives enough SMs to reach its sweet spot,
the remaining SMs are more valuable to local prefill.
Without SM partitioning, uncontrolled contention increases P99~TTFT by 19\%
and P99~TPOT by 42\% compared to our dynamic scheme, while throughput drops by 9\%.
The static 50/50 split improves over the no-partition baseline
but still wastes local prefill capacity at low batch sizes
and under-provisions encode at high batch sizes,
reducing throughput by 3\% and increasing {P99~TTFT by 11\% and} P99~TPOT by 22\% relative to our dynamic scheme.
Our dynamic scheme adapts per-batch,
approaching the latency of running encode alone at large batches
and giving the remaining SMs to local prefill at small batches.

\subsection{Arrival Pattern Sensitivity}
\label{sec:eval-sensitivity}

Our Poisson-gap dispatch rule assumes memoryless inter-arrival times.
To stress-test this assumption, we replay the end-to-end comparison (\textsc{LLaVA-34B}, $\Lambda{=}2$\,req/s) under three arrival distributions, parameterized by the coefficient of variation (CV) of inter-arrival times: Poisson (CV${=}1$), Gamma CV${=}0.5$ (more regular), and Gamma CV${=}2$ (burstier).
Table~\ref{tab:sensitivity} reports the results.

\begin{table}[t]
  \centering
  \caption{Sensitivity to arrival burstiness on \textsc{LLaVA-34B} at $\Lambda{=}2$\,req/s. CV${=}1$ corresponds to Poisson arrivals. M\,=\,mean, P\,=\,P99.}
  \label{tab:sensitivity}
  \small
  \setlength{\tabcolsep}{7pt}
  \begin{tabular}{l r r r r r}
    \toprule
    Arrival & Tput & M TTFT & P TTFT & M TPOT & P TPOT \\
    P. & (req/s) & (s) & (s) & (ms) & (ms) \\
    \midrule
    CV${=}0.5$         & 1.77 & 4.16 & \phantom{0}6.67 & 67.6 & 71.4 \\
    CV${=}1$           & 1.77 & 4.98 & 10.19 & 67.9 & 71.5 \\
    CV${=}2$           & 1.71 & 6.77 & 16.52 & 67.0 & 71.0 \\
    \bottomrule
  \end{tabular}
\end{table}

Throughput is largely insensitive to burstiness within the tested range: even under Gamma CV${=}2$, throughput drops by only 3\% (1.77$\to$1.71~req/s).
The degradation shows up mainly on tail latency: P99~TTFT increases by 62\% (10.2$\to$16.5\,s) because burst arrivals cause transient queue buildups that the Poisson-gap threshold, calibrated for the average rate, does not fully absorb.
Two factors keep this from running away: the max-age safety valve bounds the worst-case TTFT, and micro-batching still captures intra-burst requests that arrive within~$g$.
Even under our burstiest setting (CV${=}2$), the system still sustains 1.71~req/s,
showing that the end-to-end gains are not tied to perfectly memoryless arrivals.

Under more regular arrivals (CV${=}0.5$), throughput is unchanged but P99~TTFT drops by 35\% (10.2$\to$6.7\,s): more uniform inter-arrival gaps align better with the Poisson-gap threshold, reducing queue variance.
P99~TPOT remains invariant across all three distributions ($\sim$71\,ms), as expected: decode latency depends on the KV-cache access pattern, not the request arrival process.

\section{Related Work}
\label{sec:related}

\compactparagraph{Disaggregated LLM and MLLM serving.}
Disaggregating inference stages across GPU pools has become a standard technique for text-only LLM serving.
DistServe~\cite{zhong2024distserve}, Splitwise~\cite{patel2024splitwise}, 
and Mooncake~\cite{qin2024mooncake} separate prefill and decode to 
exploit their distinct compute and memory profiles.
LoongServe~\cite{loongserve2024} adds elastic sequence parallelism 
that shifts resources between the two phases at runtime, 
TPLA~\cite{tpla2026} co-designs tensor-parallel attention with disaggregation, 
and Aegaeon~\cite{aegaeon2025} pools GPUs across concurrent multi-model serving.
NVIDIA Dynamo~\cite{nvidia2025dynamo} extends disaggregation to MLLMs 
with an explicit three-stage \textsc{Encode}/\textsc{Prefill}/\textsc{Decode} pipeline, 
but executes \textsc{Encode} per-request and leaves GPU allocation to operators.
HydraInfer~\cite{hydrainfer2025} also adopts EPD disaggregation but tunes only per-stage GPU allocation, with neither \textsc{Encode}--\textsc{Prefill} offload nor joint $(B, s)$ tuning.
Parallel efforts target multimodal generative serving beyond language:
MoDM~\cite{modm2026}, TetriServe~\cite{tetriserve2026}, and DiffServe~\cite{diffserve2025}
serve diffusion and DiT workloads with their own scheduling tradeoffs orthogonal to ours.
In contrast to all of the above, our contribution is a layered architecture 
in which \textsc{Encode}-aware mechanisms and automated configuration 
are co-designed around the \textsc{Encode} stage as the single pipeline entry point.

\compactparagraph{Intra-GPU prefill--decode multiplexing.}
A recent line of work co-locates \textsc{Prefill} and \textsc{Decode} 
on the same GPU within a single LLM instance.
MuxWise~\cite{muxwise2026} uses NVIDIA Green Context~\cite{nvidia2024greencontext} 
to split SMs between the two phases with low reconfiguration overhead, 
and Bullet~\cite{bullet2026} uses \texttt{libsmctrl}~\cite{libsmctrl} 
to orchestrate dynamic spatial--temporal SM partitions for LLM serving.
Semi-PD~\cite{semipd2025} adopts MPS-based allocations with helper processes, 
Nexus~\cite{nexus2025} proposes intra-GPU prefill--decode disaggregation 
via dynamic SM partitioning, and POD-Attention~\cite{podattention2025} 
fuses the two phases at the kernel level.
These systems multiplex two dependent phases \emph{within} a single LLM instance;
our work instead multiplexes \textsc{Encode} and a local \textsc{Prefill} worker
\emph{across} the stages of a three-stage multimodal pipeline.
\textsc{Encode} co-location introduces an orthogonal offload-ratio dimension
that prefill--decode multiplexing systems do not model, 
and requires joint configuration with GPU allocation and micro-batch bound, 
a space navigated by our {\it Hybrid Auto Selection}.

\compactparagraph{GPU sharing and SM partitioning.}
CUDA MPS~\cite{nvidia2020mps} enables concurrent kernel execution
on shared GPUs with per-client resource limits and priority hints, 
but its priorities cannot preempt running kernels, 
leaving latency-sensitive workloads vulnerable to interference.
\texttt{libsmctrl}~\cite{libsmctrl} and NVIDIA Green Context~\cite{nvidia2024greencontext} 
provide strictly enforced SM partitioning with sub-microsecond overhead; 
we adopt these primitives rather than reinvent them.
LithOS~\cite{lithos2025} presents a transparent operating system for fine-grained ML multitenancy.
REEF~\cite{reef2022}, Abacus~\cite{abacus2021}, Orion~\cite{orion2024}, and GPUlet~\cite{gpulet2022} 
share GPUs among independent DNN workloads with interference-aware scheduling; 
Paella~\cite{paella2023} operates at the kernel level for fine-grained GPU sharing; 
MuxServe~\cite{muxserve2024} multiplexes multiple full LLMs on a shared GPU.
All of these co-locate \emph{independent} workloads; 
our \textsc{Encode} and local \textsc{Prefill} are \emph{dependent} 
stages of a single MLLM request, requiring pipeline-aware QoS that 
off-the-shelf GPU-sharing mechanisms do not provide.

\compactparagraph{Inference batching and scheduling.}
Orca~\cite{yu2022orca} introduced continuous batching at decode token boundaries, 
vLLM~\cite{kwon2023vllm} added PagedAttention for near-optimal KV-cache management, 
and SGLang~\cite{zheng2024sglang} improves reuse with RadixAttention.
Sarathi-Serve~\cite{agrawal2024sarathi} chunks prefill to enable stall-free batching, 
FastServe~\cite{wu2024fastserve} preempts with proactive KV management, 
AlpaServe~\cite{li2023alpaserve} exploits statistical multiplexing under bursty arrivals, 
and Niyama~\cite{niyama2025} introduces multi-class SLO-tiered scheduling.
SOLA~\cite{sola2025} optimizes SLO attainment via state-aware scheduling, 
and Klotski~\cite{klotski2025} overlaps MoE expert I/O with computation across batches.
These systems target \textsc{Prefill} or \textsc{Decode} scheduling in text-only pipelines; 
our Poisson-gap micro-batching targets the \textsc{Encode} stage specifically, 
exploiting the arrival process to derive an adaptive 
dispatch threshold for a single-shot forward pass where continuous batching does not apply.

\compactparagraph{Automated configuration for LLM serving.}
Optuna~\cite{akiba2019optuna} provides TPE-based Bayesian optimization, 
which we use as Stage~2's backbone.
Vidur~\cite{agrawal2024vidur} proposes a simulation framework for LLM configuration exploration, 
avoiding end-to-end deployment cost at the price of simulation fidelity.
ExeGPT~\cite{exegpt2024} formulates constraint-aware resource scheduling for LLM inference
as a search assuming monotonic structure over batch and parallelism choices.
Alpa~\cite{zheng2022alpha} automates parallelism-strategy search for distributed training.
Our HAS differs along two axes.
First, it combines a lightweight empirical 
proxy (per-stage capacity profiling with a bottleneck-balance score) 
for allocation screening with end-to-end benchmark measurements
for runtime-parameter refinement, avoiding both exhaustive search cost
and simulation-fidelity risk.
Second, the coupled (allocation, $B$, $s$) surface induced by our
encode-aware mechanisms is non-monotonic,
motivating TPE over the monotonic structure assumed by ExeGPT.

\section{Conclusion}
\label{sec:conclusion}

Disaggregated MLLM serving introduces a new \textsc{Encode} stage
that gates the entire pipeline,
yet existing systems run it one request at a time with operator-set GPU allocation,
which leaves substantial GPU capacity idle
and forces operators to navigate a configuration space too large to tune by hand.
To address this, we propose \projectname, a layered serving framework
that repositions \textsc{Encode} as the pipeline's control point.
\projectname combines an offline configuration search,
which jointly tunes GPU allocation, encode batch bound, and prefill offload ratio,
with three encode-aware runtime mechanisms
that expose those parameters and enforce the latency guarantees
needed for the search to be reliable.
Across image, video, and audio MLLM workloads on an 8-GPU server,
\projectname delivers up to $4.3\times$ higher goodput than strong production baselines
while reclaiming most of the encode-side capacity that prior systems leave unused.
Looking ahead, we plan to validate the design at multi-node datacenter scale,
where cross-node embedding transfer and HAS's search budget under thousands of candidate
configurations both warrant dedicated study,
and to extend the dispatch rule with online arrival-rate estimation for various workloads.

\begin{acks}
\revisioncr{
The authors want to extend their appreciation to all the anonymous reviewers for their valuable and thorough feedback. 
All of these constructive suggestions have greatly contributed to enhancing this paper. 
This work was supported in part by the National Science Foundation (NSF) under the awards of 
CCF-2428108, CCF-2333895, OAC-2403090, and CSR-2341378.
Any errors and opinions are not those of the NSF and are attributable solely to the author(s). 
}
\end{acks}

\bibliographystyle{ACM-Reference-Format}
\bibliography{bib/reference}

\end{document}